\documentclass[11pt,twoside]{article}

\usepackage{asp2006}
\usepackage{epsf}
\usepackage{graphicx}

\begin{document}
\title{GRO J1655-40: from ASCA and XMM-Newton Observations}   
\author{Xiao-Ling~Zhang\altaffilmark{1},
  Shuang~Nan~Zhang\altaffilmark{2,3,4},
  Gloria~Sala\altaffilmark{1},
  Jochen~Greiner\altaffilmark{1}, 
  Yuxin~Feng\altaffilmark{3,4},
  Yangsen~Yao\altaffilmark{5}}
\altaffiltext{1}{MPE, Postfach 1312, 85741 Garching, Germany, zhangx@mpe.mpg.de}
\altaffiltext{2}{Tsinghua Univ, 100084, Beijing, China}
\altaffiltext{3}{U. of Alabama, Huntsville, AL 35899, USA}
\altaffiltext{4}{NSSTC, Sparkman Dr. 320, Huntsville AL 35805, USA}
\altaffiltext{5}{MIT Kavli Inst. for Astro. and Space Research, 70 Vassar Street, Cambridge, MA 02139}

\begin{abstract} 
We have analysed four ASCA observations (1994--1995, 1996--1997)
and three XMM-Newton observations (2005) of this source,
in all of which the source is in high/soft state.
We modeled the continuum spectra with relativistic disk
model {\em kerrbb}, estimated the spin of the central black hole,
and constrained the spectral hardening factor $f_{\mathrm{col}}$
and the distance.
If {\em kerrbb} model applies, for normally used value 
of $f_{\mathrm{col}}$ (1.7), the distance cannot be very small,
and $f_{\mathrm{col}}$ changes with observations.
\end{abstract}

\noindent
{\bf 1. Background} 
\vspace{0.1in} 
\\
GRO~J1655-40, the second microquasar (after GRS 1915-105), 
had X-ray outbursts in 1994-1995, 1996-1997, 2005. 
Its geometric parameters are 
considered 
best determined:
mass $ M_{\mathrm{BH}} = 7.0 \pm 0.2M_{\odot} $, 
inclination angle $ \theta = 69.50\deg \pm 0.08$ \citep{orosz1997}, 
distance $ D = 3.2 \pm 0.2$~kpc \citep{hjellming1995}, 
which makes it a very good laboratory of studying black holes and environments.

The spin of the central black hole has been estimated by various authors with 
various methods \citep[see, e.g.,][]{zhang_spin, abramowicz2001,
aschenbach2004, shafee2006}, and the reported value range
 from 0.2 \citep{abramowicz2001} to 0.996 \citep{aschenbach2004}.

In estimating black hole spin from continuum spectral modeling, 
the color correction factor
$ f_{\mathrm{col}} = T_{\mathrm{col}} / T_{\mathrm{eff}}$,
is one of the key factors. The normally used value of 
$ f_{\mathrm{col}} $ is 1.7, following \citet{shimura1995},
while many authors believe it should not be constant 
\citep[see, e.g.,][]{merloni2000}.
The distance is also very important. The widely accepted value 
$3.2 \pm 0.2$~kpc
was challenged by \citet{foellmi2006}, who gave an upper limit of 1.7~kpc. \\

\noindent
{\bf 2. Observations, data reduction and model fitting} 
\vspace{0.1in} 
\\
We analysed three ASCA observations during the 1994--1995 and 
the 1996--1997 outbursts, and three XMM-Newton observations 
during the 2005 outburst, in all of which
the source was in high/soft state.
For ASCA, only GIS2 data were used, after gain correction and
deadtime correction.
For XMM-Newton, only Epic-pn data were used, 
after correction for rate-dependent 
Charge-Transfer-Efficiency \citep{sala2006}.

The classical way of estimating black hole spin from the continuum 
spectral fitting is to fit the spectra with disk models, and obtain 
the spin directly or indirectly. All models take the source distance
as parameter, and most models treat the disk as multi-temperature 
black-body rings and the derived spin value depends on the apparent/effective
temperature ratio. 

\begin{figure}[!ht]
\begin{center}
\includegraphics[width=0.7\textwidth]{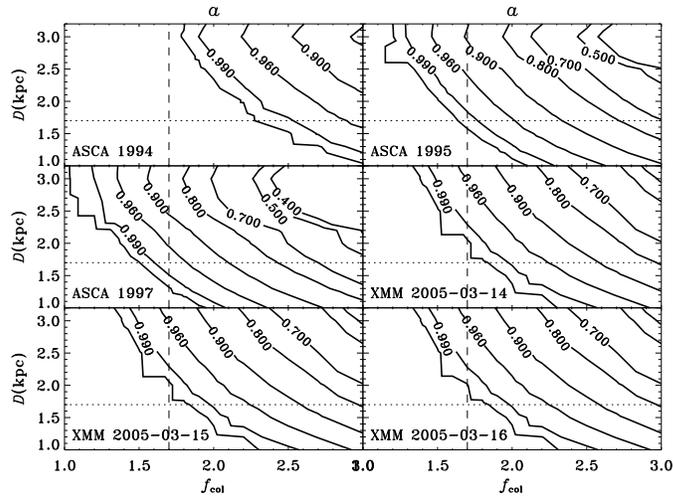}
\caption{Contour of black hole specific angular momentum $a$ versus 
distance $D$ and $f_{\mathrm{col}}$.
The two dotted lines indicate D=1.7kpc, and $f_{\mathrm{col}}=1.7$.
}
\label{fig:spin}
\end{center}
\end{figure}

The relativistic disk 
model {\em kerrbb} in {\em XSPEC} was used in the fitting. 
We let $f_{\mathrm{col}}$ vary from 1.0 to 3.0, and
$D$ vary from 1.0~kpc and 3.2~kpc. 
For each combination of $f_{\mathrm{col}}$ and D, 
we fitted the data sets and obtained a spin value, if the fit was acceptable 
($\chi^2/dof<2$). 
The contour of the derived spin $a$ over $D$ and $f_{\mathrm{col}}$ 
are shown in the Fig.~\ref{fig:spin}. \\

\noindent
{\bf 3. Conclusion} \vspace{0.1in} 
\\
From Fig.~\ref{fig:spin} we can see, 
1. for the normally used $f_{\mathrm{col}}$ value 1.7, {\em kerrbb} model does 
not favor small distance; 
2. because the black hole spin and the source distance should be constant,
$f_{\mathrm{col}}$ changes dramatically between these observations.

\end{document}